\documentclass[10pt]{article}
\input epsf

\begin{document}

\title{The thermal and kinematic Sunyaev-Zel'dovich effects revisited}
\author{A. Sandoval-Villalbazo$^a$ and L.S. Garc\'{\i}a-Col\'{\i}n$^{b,\,c}$ \\
$^a$ Departamento de F\'{\i}sica y Matem\'{a}ticas, Universidad Iberoamericana \\
Lomas de Santa Fe 01210 M\'{e}xico D.F., M\'{e}xico \\
E-Mail: alfredo.sandoval@uia.mx \\
$^b$ Departamento de F\'{\i}sica, Universidad Aut\'{o}noma Metropolitana \\
M\'{e}xico D.F., 09340 M\'{e}xico \\
$^c$ El Colegio Nacional, Centro Hist\'{o}rico 06020 \\
M\'{e}xico D.F., M\'{e}xico \\
E-Mail: lgcs@xanum.uam.mx} \maketitle
\bigskip
\bigskip
\begin{abstract}
{\small This paper shows that a simple convolution integral
expression based on the mean value of the isotropic frequency
distribution corresponding to photon scattering off electrons
leads to useful analytical expressions describing the thermal
Sunyaev-Zel'dovich effect. The approach, to first order in the
Compton parameter is able to reproduce the Kompaneets equation
describing the effect. Second order effects in the parameter
$z=\frac{kT_{e}}{mc^{2}}$ induce a slight increase in the
crossover frequency.}
\end{abstract}

\section{\textbf{Introduction}}

Ever since R. Sunyaev and Ya. B. Zel'dovich predicted the
distortion of the spectral density of the Cosmic Microwave
Background Radiation (CMBR) in 1969  \cite{two} \cite{three} by
the ionized plasma in globular clusters, whose main constituent is
a relatively dense electron gas, dozens of papers have been
published addressing various aspects of this phenomenon. This
avalanche of works is mainly due to the undeniable importance of
such effect, not only because it yields information on deviations
from the isotropy and homogeneity of the Universe, but also
provides an aid in determining other important cosmological
parameters such as the baryonic density of matter, Hubble's
constant, age and velocity of massive clusters, and others. These
features have been thoroughly reviewed in recent literature so we
shall not dwell with them here \cite{four}-\cite{six}.

Although practically every cosmologist or astrophysicist believes
that this effect is now ``clearly'' understood, there are reasons
that we believe will help clarifying some of the still subtle
details that remain unclear in the available treatments. Firstly,
the early attempts to study this effect by visualizing the motion
of the photons through the hot electron gas in the cluster as a
diffusion process which, in the non-relativistic limit is
described by the famous Kompaneets equation \cite{seven}
\cite{Weinman}. The analysis of the results obtained through the
hot electron gas in the cluster through the use of this equation
are well-known \cite{four}-\cite{five}. Nevertheless, it was soon
realized by many authors, including Sunyaev himself, that due to
the very small probability that an incoming photon is scattered
even once in its passage through the gas, a Compton-like
scattering of a photon off an electron was a much more suitable
mechanism to explain the effect. This trend of ideas has been
recently reviewed by Dolgov et al \cite{Steen} where the reader
may find more literature about the subject. The somewhat
intriguing result is that both approaches lead to identical
results, a fact that needs further clarification. We address this
question in this paper.

Secondly, the role of the optical depth $\tau$, related to the
distance a photon can travel in the plasma before it is scattered
off an electron, plays in the way the broadening of the spectral
lines of the scattered photon is modified by the Doppler effect,
has not yet been justified in a convincing manner. Here we address
this question form an entirely different point of view. And
thirdly, we also show that this different interpretation of $\tau$
also leads to the kinematic Sunyaev-Zel'dovich (SZ) effect in a
rather trivial way.

\section{\textbf{The method}}
We begin our discussion by establishing a general expression for
the full distorted spectrum, $I(\nu)$ of the scattered radiation
off the plasma. Let $\tau$ be the optical depth as discussed
above. This quantity is a direct measure of the probability that a
photon is scattered off an electron in the gas. Thus, since the
effect is small, $(1-\tau)$ is the probability that the photon
traverses the plasma unscattered. Therefore,
\begin{equation}
\begin{array}{c}
I\left( \nu \right) =(1-\tau )\int_{-\infty}^{\infty }I_{o}\left(
\bar{\nu}\right)
\delta (\bar{\nu}-\nu )\,d\bar{\nu}+ \\
\;\;\;\;\;\frac{\tau }{\sqrt{2\pi }\sigma \left( \nu \right)
}\int_{-\infty }^{\infty }I_{o}\left( \bar{\nu}\right) \exp \left[
-\left( \frac{\bar{\nu} -(1-2\,z)\,\nu }{2\sigma \left( \nu
\right) }\right) ^{2}\right] \,d\bar{\nu}
\end{array}
\label{uno}
\end{equation}
The first term in Eq. ( \ref{uno}) is self-explanatory, the lower
limit justified since $I_{o}(\overline{\nu})$, the incoming flux
is zero for negative frequencies. The second term, the important
one, defines a Gaussian probability function describing scattering
of a photon with incoming frequency $\overline{\nu}$ off an
electron whose average kinetic energy in the gas, in the
non-relativistic approximation is $ \frac{1}{2}k T_{e}$. This
function has been thoroughly discussed in earlier work \cite{one}
so we shall not discuss it here further. In Eq.( \ref{uno}) the
incoming flux is given, for $\overline{\nu}>0$ by:
\begin{equation}
I_{o}(\overline{\nu})=\frac{2h \overline{\nu}^{3}}{c^{2}}
\frac{1}{e^{\frac{h \overline{\nu}}{k T}}-1} \label{dos}
\end{equation}
where $T\equiv2.726\,\,K$, the temperature of the cosmic
background radiation, $k$ is Boltzmann's constant, $h$ is Planck's
constant, and $c$ the speed of light. Further, $\sigma \left( \nu
\right) $ is the width of the spectral line at frequency $\nu $
and its squared value reads \cite{one}:
\begin{equation}
\sigma ^{2}\left( \nu \right) =2\,\frac{kT_{e}}{m\,c^{2}}\nu ^{2}=2\,z\,\nu
^{2}  \label{tres}
\end{equation}
In Eq.(\ref{dos}) $T_{e}$ is the temperature of the electron
scatterers, $m$ is the electron mass and $I_{o}\left(
\bar{\nu}\right) \equiv 0$ for $\bar{\nu}<0$.

Notice should be made of the fact that the mathematical expression
for $\sigma ^{2}(\nu)$ corresponds to what in the literature is
called the inverse Compton scattering \cite {Lightman}. In
particular, this is the reason of the factor $ 2z$ in such
equation. To show this fact we take into account the frequency
distribution function for photons with incoming frequency $\tilde{
\nu}$ isotropically scattered by electrons with velocities
$u=\beta \,c$\cite {Lightman},
\begin{equation}
P_{iso}\left( \nu ,\tilde{\nu},\beta \right) =\frac{\tilde{\nu}}{(2\beta
\gamma )^{2}\nu }\left\{
\begin{array}{c}
\left( 1+\beta \right) \frac{\tilde{\nu}}{\nu }-1+\beta
,\;\frac{1-\beta }{
1+\beta }\leq \frac{\tilde{\nu}}{\nu }\leq 1 \\
1+\beta -\left( 1-\beta \right) \frac{\tilde{\nu}}{\nu
},\;1<\frac{\tilde{\nu }}{\nu }\leq \frac{1+\beta }{1-\beta }
\end{array}
\right.   \label{tres_cinco}
\end{equation}
where $\gamma =\left( 1-\beta^{2} \right) ^{-1/2}$. The expected
value for the photon frequency at fixed $\beta $ is then given by:
\begin{equation}
\left\langle \nu \right\rangle =\frac{\int_{\frac{1+\beta
}{1-\beta }}^{ \frac{1-\beta }{1+\beta }}\nu P_{iso}\left( \nu
,\tilde{\nu},\beta \right) d \tilde{\nu}}{\int_{\frac{1+\beta
}{1-\beta }}^{\frac{1-\beta }{1+\beta } }P_{iso}\left( \nu
,\tilde{\nu},\beta \right) d\tilde{\nu}}=-\frac{1+\beta
^{2}}{-1+\beta ^{2}}\nu \cong \left( 1+2\beta ^{2}\right) \nu
\label{tres_75}
\end{equation}
Direct application of the equipartition energy theorem allows us to identify
\newline
$<$ $\beta
^{2}>=<\frac{mu^{2}}{2}>\frac{2}{mc^{2}}=\frac{kT_{e}}{2}\frac{2}{
mc^{2}}=z$. Physically, this implies that a typical photon
scattered off an electron will have a temperature dependent blue
shift simply given by $2z\nu ^{2}$. It is interesting to notice
that a similar ``frequency shift approach'' leads to the
analytical well known expression of the kinematic SZ effect. This
is shown in the appendix.

Eq. (\ref{uno}) can be evaluated analytically  leading to an
expression which is identical to the one reported in the
literature. Setting in Eq. ( \ref{uno})
\begin{equation}
\alpha =\frac{\bar{\nu}-(1-2z)\,\nu }{\sigma \left( \nu \right) }
\label{cuatro}
\end{equation}
we get that
\begin{equation}
I\left( \nu \right) -I_{o}\left( \nu \right) =-\tau I_{o}\left(
\nu \right) + \frac{1}{\sqrt{\pi }}\int_{-\infty }^{\infty
}I_{o}\left( \nu +\Delta \nu \left( \alpha \right) \right) \exp
\left[ -\alpha ^{2}\right] \,d\alpha \label{cinco}
\end{equation}
where
\begin{equation}
\Delta \nu \left( \alpha \right) =-2\,z\nu +2z^{1/2}\nu \alpha   \label{seis}
\end{equation}
Since $z$ is a small number \cite{four} and $I_{o}$ is an
analytical function of $\nu $ we may expand the integrand in Eq.
(\ref{cinco}) in a Taylor series:
\begin{equation}
I_{o}\left( \nu +\Delta \nu \left( \alpha \right) \right) =I_{o}\left( \nu
\right) +\Delta \nu \left( \alpha \right) \frac{\partial I_{o}}{\partial \nu
}+...  \label{siete}
\end{equation}
All odd powers of $\alpha $ yield zero upon integration so, after a
straightforward calculation, we get that
\begin{equation}
I\left( \nu \right) -I_{o}\left( \nu \right) =-2y\,\nu
\frac{\partial I_{o}}{
\partial \nu }+y\,\nu ^{2}\frac{\partial ^{2}I_{o}}{\partial \nu ^{2}}+2\tau
z^{2}\nu ^{2}\frac{\partial ^{2}I_{o}}{\partial \nu ^{2}}-2\tau
z^{2}\nu ^{3} \frac{\partial ^{3}I_{o}}{\partial \nu
^{3}}+\frac{\tau z^{2}\nu ^{4}}{2} \frac{\partial
^{4}I_{o}}{\partial \nu ^{4}}  \label{ocho}
\end{equation}
To first order in $y=\,\tau $ $z$\ we reproduce \emph{exactly} the analytic
expression for the distortion curve \cite{two}-\cite{three}. It is also
interesting to notice that, taking into account the second order terms in $z$
included in Eq. (\ref{ocho}), the crossover frequency $\nu _{c}$
corresponding to the solution of equation
\begin{equation}
I\left( \nu \right) -I_{o}\left( \nu \right) =0  \label{nueve}
\end{equation}
is now clearly dependent on the value of $z$ namely, on the
electron temperature, a feature not ordinarily recognized for a
non-relativistic intra-cluster gas. For typical values of $\tau
=10^{-2}$, and $kT_{e}=3KeV$ a numerical calculation yields $\nu
_{c}=221.34GHz$, while the value obtained by considering only
linear terms in $z$ yields $\nu _{c}=217.16$ GHz, a difference of
nearly $ 1.74\%$. Fig. 1 shows the plot of the analytical
non-relativistic expression up to second order in $z^{2}$,
compared with its first order counterpart for a temperature of
$kT=3 KeV$.  It is interesting to notice that, although a
\emph{non-relativistic} Maxwellian was used in the convolution
integral, the  Wien side of the distortion curve is practically
identical to its relativistic counterpart, as presented in Ref.
\cite{Sazonov}. This remark is confirmed in Fig. 2 with $kT=15
KeV$.

\begin{figure}
\epsfxsize=3.4in \epsfysize=2.6in
\epsffile{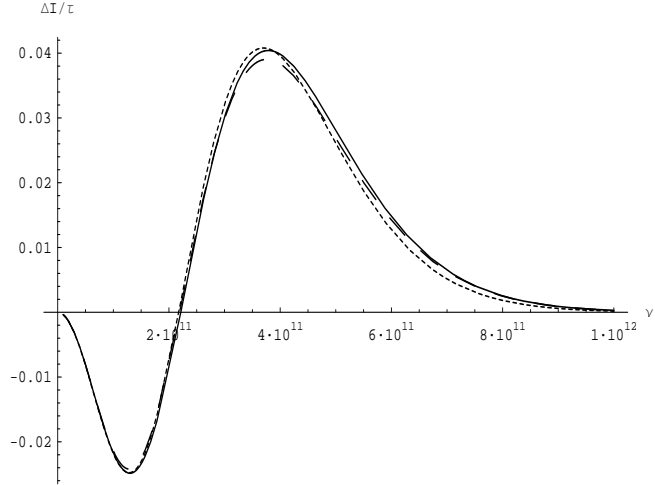}\vspace{0.5cm} \caption{CMBR SZ effect
distortion $\frac{\Delta I}{\tau}$ as computed by Eq.(10), with
$kT=3 KeV$ (solid curve). The intensity change is measured in
units of $\frac{(hc)^{2}}{2(kT_{o})^{3}}$ and the frequency is
given in Hz. The short-dashed line next to it shows the intensity
change corresponding to the Kompaneets-based expression. The
thermal relativistic SZ effect is represented by the long-dashed
curve}
\end{figure}
\vspace{0.5cm}

The result expressed in Eq. (\ref{ocho}) is of significant
importance. Eq. ( \ref{ocho}) contains the famous Kompaneets
equation \cite{five}-\cite{Weinman} describing the thermal SZ
effect using a diffusive mechanism to account for the motion of
photons in an optically thin electron gas where the electron
temperature $T_{e}>>T$, the photon temperature (see specifically
Ref. \cite{six}). One must notice that, since most photons
\emph{are not scattered even once}, a diffusion approximation
would then hardly seem adequate \cite{four}. Nevertheless, the
diffusion mechanism is mathematically identical to the
absortion-emission process of photons by electrons to first order
in $z$. The problem posed in Ref. \cite{one} requiring an
explanation of why two different mechanisms lead to the same
result is now mathematically solved. Finally, we want to stress
that it is now clear why the curves describing the thermal SZ
effect obtained by numerical integration of different mathematical
expressions turn out to be identical. We therefore expect that the
relativistic SZ effect and other photon scattering problems in hot
plasmas can be handled by a similar procedure.

\begin{figure}
\epsfxsize=3.4in \epsfysize=2.6in
\epsffile{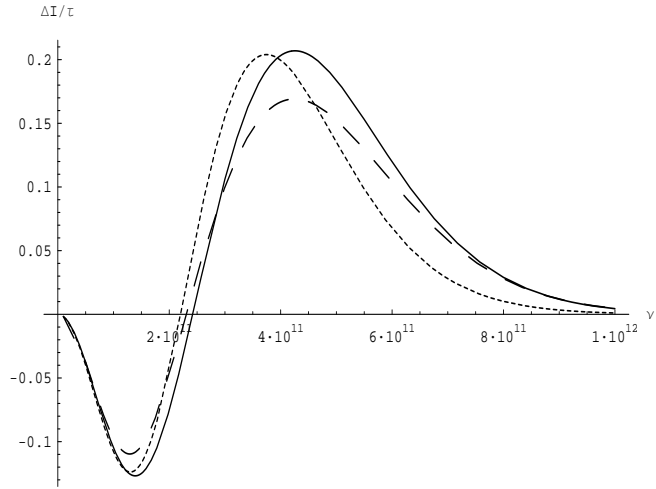}\vspace{0.5cm} \caption {The same as in Fig.
1, with $kT=15 KeV$ }
\end{figure}
\vspace{0.5cm}

\bigskip

This work has been supported by CONACyT (M\'{e}xico), project 41081-F.

\bigskip

\textbf{Appendix}

\medskip

The central idea of the convolution integral approach to CMBR distortions is
that, in a dilute gas, a scattering law $G(\bar{\nu},\nu )$ is given by what
in statistical physics is known as the \emph{\ dynamic structure factor},
where $\bar{\nu}$ and $\nu $ are the corresponding incoming/outcoming photon
frequencies. Then, one computes the distorted occupation number through the
convolution integral
\[
n\left( \nu \right) =\int_{0}^{\infty }n_{o}\left(
\bar{\nu}\right) G(\bar{ \nu},\nu )\,d\bar{\nu}
\]
with
\[
n_{o}\left( \bar{\nu}\right)
=\frac{1}{e^{\frac{h\bar{\nu}}{kT}}-1}
\]
In the non-relativistic kinetic SZ effect, one can easily show
that a suitable structure factor is given by:

\[
G_{k}(\bar{\nu},\nu )\,=(1-\tau )\,\delta (\bar{\nu}-\nu )+\tau
\,\delta ( \bar{\nu}-(1-\frac{U\,}{c})\nu )
\]
$U$ being the cluster velocity (\emph{approaching the observer}).
The optical depth $\tau $ measures essentially the proportion of
photons being captured (scattered) by electrons. The intergalactic
gas cloud in clusters of galaxies has an optical depth $\tau \sim
10^{-2}$ .

The formalism basically states that a fraction $(1-\tau )\,\ $ of
photons passes through the cluster without being scattered, while
a $\tau $ fraction of them shifts its frequency through the
ordinary Doppler effect. The integrals can be evaluated leading to
an expression which is identical to the one reported in the
literature. Indeed, since
\[
n\left( \nu \right) =(1-\tau )\int_{0}^{\infty }n_{o}\left(
\bar{\nu}\right) \,\delta (\bar{\nu}-\nu )d\bar{\nu}+\tau
\,\int_{0}^{\infty }n_{o}\left( \bar{\nu}\right) \delta
(\bar{\nu}-(1-\frac{U\,}{c})\nu )d\bar{\nu}
\]
then
\[
n\left( \nu \right) =n_{o}\left( \nu \right) (1-\tau )\,+\tau
\,\int_{0}^{\infty }n_{o}\left( \bar{\nu}\right) \delta
(\bar{\nu}-(1-\frac{ U\,}{c})\nu )d\bar{\nu}
\]
If we now perform the substitution
\[
\hat{\alpha}=\bar{\nu}-(1-\frac{U\,}{c})\nu
\]
we may expand the integrand in a Taylor series. Straightforward
calculation leads to the ordinary expression for the kinetic SZ
effect:
\[
n\left( \nu \right) -n_{o}\left( \nu \right) =-\frac{U\,\tau
}{c}\,\frac{
\partial n_{o}}{\partial \nu }
\]
If we now define $I_{k}\left( \nu \right) $ as the distorted
kinetimatic SZ spectrum and consider the expressions $x=\frac{h\nu
}{kT}$ and $\frac{ n\left( \nu \right) -n_{o}\left( \nu \right)
}{n_{o}\left( \nu \right) }= \frac{I_{k}\left( \nu \right)
-I_{o}\left( \nu \right) }{I_{o}\left( \nu \right) }$, we finally
get the desired expression:
\[
I_{k}\left( \nu \right) -I_{o}\left( \nu \right)
=\frac{2(kT)^{3}}{\left( hc\right) ^{2}}\frac{x^{4}e^{x}}{\left(
e^{x}-1\right) ^{2}}\frac{U\,\tau }{c }
\]

\end{document}